\documentclass[cits]{JINST}

\title{Tests of a Particle Flow Algorithm with CALICE Test Beam Data}

\author{\centering 
\LARGE\bf The CALICE Collaboration
}

\author{\centering
C.\,Adloff, 
J.\,Blaha, 
J.-J.\,Blaising, 
C.\,Drancourt,
A.\,Espargili\`{e}re, 
R.\,Gaglione, 
N.\,Geffroy, 
Y.\,Karyotakis, 
J.\,Prast,
G.\,Vouters
\\ \it
Laboratoire d'Annecy-le-Vieux de Physique des Particules, Universit\'{e} de Savoie,
CNRS/IN2P3,
9 Chemin de Bellevue BP110, F-74941 Annecy-le-Vieux CEDEX, France
}

\author{\centering
K.\,Francis,
J.\,Repond, 
J.\,Smith\footnote{Also at University of Texas, Arlington},
L.\,Xia 
\\ \it
Argonne National Laboratory,
9700 S.\ Cass Avenue,
Argonne, IL 60439-4815,
USA}

\author{\centering
E.\,Baldolemar, 
J.\,Li\footnote{Deceased}, 
S.\,T.\,Park, 
M.\,Sosebee, 
A.\,P.\,White, 
J.\,Yu
\\ \it
Department of Physics, SH108, University of Texas, Arlington, TX 76019, USA
}

\author{\centering
T.\,Buanes, G.\,Eigen
\\ \it
University of Bergen, Inst.\, of Physics, Allegaten 55, N-5007 Bergen, Norway
}

\author{\centering
Y.\,Mikami, 
N.\,K.\,Watson 
\\ \it
University of Birmingham,
School of Physics and Astronomy,
Edgbaston, Birmingham B15 2TT, UK
}

\author{\centering 
T.\,Goto, 
G.\,Mavromanolakis\footnote{Now at CERN}, 
M.\,A.\,Thomson, 
D.\,R.\,Ward, 
W.\,Yan\footnote{Now at Dept.\, of Modern Physics, Univ. of Science and Technology of China, 96 Jinzhai Road, Hefei, Anhui, 230026, P.\, R.\, China}
\\ \it
University of Cambridge, Cavendish Laboratory, J J Thomson Avenue, CB3 0HE, UK
}

\author{\centering 
D.\,Benchekroun, 
A.\,Hoummada, 
Y.\,Khoulaki
\\ \it
Universit\'{e} Hassan II A\"{\i}n Chock, Facult\'{e} des sciences.\, B.P. 5366 Maarif, Casablanca, Morocco
}

\author{\centering
M.\,Benyamna, 
C.\,C\^{a}rloganu, 
F.\,Fehr, 
P.\,Gay, 
S.\,Manen, 
L.\,Royer
\\ \it
Clermont Univertsit\'e, Universit\'e Blaise Pascal, CNRS/IN2P3, LPC, BP
10448, F-63000 Clermont-Ferrand, France
}

\author{\centering
G.\,C.\,Blazey,
A.\,Dyshkant, 
J.\,G.\,R.\,Lima, 
V.\,Zutshi
\\ \it
NICADD, Northern  Illinois University,
Department of Physics,
DeKalb, IL 60115,
USA
}

\author{\centering 
J.\,-Y.\,Hostachy, 
L.\,Morin
\\ \it
Laboratoire de Physique Subatomique et de Cosmologie - Universit\'{e} Joseph Fourier Grenoble 1 -
CNRS/IN2P3 - Institut Polytechnique de Grenoble,
53, rue des Martyrs,
38026 Grenoble CEDEX, France
}

\author{\centering 
U.\,Cornett, 
D.\,David, 
R.\,Fabbri, 
G.\,Falley, 
K.\,Gadow, 
E.\,Garutti,
P.\,G\"{o}ttlicher, 
C.\,G\"{u}nter,
S.\,Karstensen, 
F.\,Krivan,
A.\,-I.\,Lucaci-Timoce\footnotemark[3], 
S.\,Lu, 
B.\,Lutz, 
I.\,Marchesini, 
N.\,Meyer,
S.\,Morozov, 
V.\,Morgunov\footnote{On leave from ITEP}, 
M.\,Reinecke, 
F.\,Sefkow, 
P.\,Smirnov,
M.\,Terwort,
A.\,Vargas-Trevino, 
N.\,Wattimena, 
O.\,Wendt
\\ \it
DESY, Notkestrasse 85,
D-22603 Hamburg, Germany
}

\author{\centering  
N.\,Feege, 
J.\,Haller, 
S.\,Richter, 
J.\,Samson
\\ \it
Univ. Hamburg,
Physics Department,
Institut f\"ur Experimentalphysik,
Luruper Chaussee 149,
22761 Hamburg, Germany
}

\author{\centering 
P.\,Eckert,
A.\,Kaplan,
 H.\,-Ch.\,Schultz-Coulon,
 W.\,Shen,
 R.\,Stamen,
 A.\,Tadday
\\ \it
 University of Heidelberg, Fakultat fur Physik und Astronomie,
Albert Uberle Str. 3-5 2.OG Ost,
D-69120 Heidelberg, Germany
}

\author{\centering 
B.\,Bilki, E.\,Norbeck, 
Y.\,Onel
\\ \it
University of Iowa, Dept. of Physics and Astronomy,
203 Van Allen Hall, Iowa City, IA 52242-1479, USA
}

\author{\centering 
G.\,W.\,Wilson
\\ \it
University of Kansas, Department of Physics and Astronomy,
Malott Hall, 1251 Wescoe Hall Drive, Lawrence, KS 66045-7582, USA
}

\author{\centering 
K.\,Kawagoe,  
S.\,Uozumi\footnote{Now at Kyungpook National University, South Korea.}
\\ \it
 Department of Physics, Kobe University, Kobe, 657-8501, Japan
}

\author{\centering 
P.\,D.\,Dauncey, 
A.\,-M.\,Magnan
\\ \it
Imperial College London, Blackett Laboratory,
Department of Physics,
Prince Consort Road,
London SW7 2AZ, UK 
}

\author{\centering 
M.\,Wing
\\ \it
Department of Physics and Astronomy, University College London,
Gower Street,
London WC1E 6BT, UK
}

\author{\centering 
F.\,Salvatore\footnote{Now at University of Sussex, Physics and Astronomy Department, Brighton, Sussex, BN1 9QH, UK}
\\ \it
Royal Holloway University of London,
Dept. of Physics,
Egham, Surrey TW20 0EX, UK
}

\author{\centering 
E.\,Calvo~Alamillo, 
M.-C.\, Fouz, 
J.\,Puerta-Pelayo 
\\ \it
CIEMAT, Centro de Investigaciones Energeticas, Medioambientales y Tecnologicas, Madrid, Spain 
}

\author{\centering 
V.\,Balagura, 
B.\,Bobchenko, 
M.\,Chadeeva, 
M.\,Danilov, 
A.\,Epifantsev, 
O.\,Markin$^\spadesuit$,
R.\,Mizuk, 
E.\,Novikov, 
V.\,Rusinov, 
E.\,Tarkovsky
\\ \it
Institute of Theoretical and Experimental Physics, B. Cheremushkinskaya ul. 25,
RU-117218 Moscow, Russia
}

\author{\centering 
N.\,Kirikova,
V.\,Kozlov, 
Y.\,Soloviev 
\\ \it
P.\,N.\, Lebedev Physical Institute,
Russian Academy of Sciences,
117924 GSP-1 Moscow, B-333, Russia
}

\author{\centering 
P.\,Buzhan, B.\,Dolgoshein, A.\,Ilyin, V.\,Kantserov, V.\,Kaplin, A.\,Karakash, E.\,Popova, S.\,Smirnov 
\\ \it
Moscow Physical Engineering Inst., MEPhI,
Dept. of Physics,
31, Kashirskoye shosse,
115409 Moscow, Russia
}

\author{\centering 
A.\,Frey\footnote{Now at University of G\"{o}ttingen, II. Physikalisches Institut
Friedrich-Hund-Platz 1, 37077 G\"ottingen, Germany}, 
C.\,Kiesling,
K.\,Seidel, 
F.\,Simon,
C.\,Soldner, 
L.\,Weuste
\\ \it
Max Planck Inst. f\"ur Physik,
F\"ohringer Ring 6,
D-80805 Munich, Germany
}

\author{\centering 
J.\,Bonis, 
B.\,Bouquet,    
S.\,Callier, 
P.\,Cornebise, 
Ph.\,Doublet,
F.\,Dulucq, 
M.\,Faucci Giannelli, 
J.\,Fleury,
H.\,Li\footnote{Now at LPSC Grenoble},  
G.\,Martin-Chassard, 
F.\,Richard, 
Ch.\,de la Taille, 
R.\,P\"{o}schl, 
L.\,Raux,  
N.\,Seguin-Moreau, 
F.\,Wicek
\\ \it

Laboratoire de l'Acc\'{e}l\'{e}rateur Lin\'{e}aire, Centre
Scientifique d'Orsay, Universit\'{e} de Paris-Sud XI, CNRS/IN2P3, BP
34, B\^atiment 200, F-91898 Orsay CEDEX, France
}

\author{\centering 
M.\,Anduze,
V.\,Boudry, 
J-C.\,Brient, 
D.\,Jeans, 
P.\,Mora de Freitas, 
G.\,Musat, 
M.\,Reinhard, 
M.\,Ruan,  
H.\,Videau
\\ \it
 Laboratoire Leprince-Ringuet (LLR)  -- \'{E}cole Polytechnique, CNRS/IN2P3, F-91128 Palaiseau, France
}

\author{\centering 
B.\,Bulanek,
J.\,Zacek 
\\ \it
Charles University, Institute of Particle \& Nuclear Physics,
V Holesovickach 2,
CZ-18000 Prague 8, Czech Republic  
}

\author{\centering 
J.\,Cvach, 
P.\,Gallus, 
M.\,Havranek, 
M.\,Janata, 
J.\,Kvasnicka,
D.\,Lednicky,
M.\,Marcisovsky, 
I.\,Polak, 
J.\,Popule, 
L.\,Tomasek, 
M.\,Tomasek, 
P.\,Ruzicka, 
P.\,Sicho, 
J.\,Smolik, 
V.\,Vrba, 
J.\,Zalesak 
\\ \it
Institute of Physics, Academy of Sciences of the Czech Republic, Na Slovance 2,
CZ-18221 Prague 8, Czech Republic
}

\author{\centering 
B.\,Belhorma,
H.\,Ghazlane
\\ \it
Centre National de l'Energie, des Sciences et des Techniques Nucl\'{e}aires, 
B.P. 1382, R.P. 10001, Rabat, Morocco
}

\author{{\centering              
T.\,Takeshita
\\ \it
Shinshu Univ.\,,
Dept. of Physics,
3-1-1 Asaki,
Matsumoto-shi, Nagano 390-861,
Japan \\
}
\it
$^\spadesuit$ Corresponding author\newline
E-mail: \email{markin@itep.ru}

}

\abstract{
The studies presented in this paper provide a first experimental test of the Particle Flow Algorithm (PFA) concept using data recorded in high granularity calorimeters.  Pairs of overlaid pion showers from CALICE 2007 test beam data are reconstructed by the PandoraPFA program developed to implement PFA for a future lepton collider. Recovery of a neutral hadron's energy in the vicinity of a charged hadron is studied. The impact of the two overlapping hadron showers on energy resolution is  investigated.  The dependence of the confusion error on the distance between a 10~GeV neutral hadron and a charged pion is derived for pion energies of 10 and 30~GeV which are representative of a 100~GeV jet. The comparison of these test beam data results with Monte Carlo simulation is done for various hadron shower models within the GEANT4 framework.  The results for simulated particles and for beam data  are in good agreement thereby providing support for previous simulation studies of the power of Particle Flow Calorimetry at a future lepton collider.}


\begin{document}

\section{Introduction}
The experimental program for the future International Linear Collider (ILC)~\cite{ilc} and Compact Linear Collider (CLIC)~\cite{clic} assumes a particle detector with unprecedented energy resolution for jets, about a factor two to three better than was achieved at LEP. Such a high resolution is crucial in investigations of Higgs boson properties and potentially decisive in searching for the lightest SUSY particles as well as in studies of Strong Electroweak Symmetry Breaking, where it would allow $W$ and $Z$ hadronic decays to be distinguished~\cite{ilc,TDR}. 

The most advanced and promising way to reach such a resolution utilizes the concept of a Particle Flow Algorithm (PFA).  Using PFAs, ideally only the energy of neutral particles is measured in the calorimeters, while the charged particle energy is reconstructed in a tracker where the resolution is much better. Since the majority of particles in jets are charged and therefore can be identified in the tracker, the PFA approach outperforms the traditional calorimetric approach which derives the energy of the whole jet from calorimetric measurements. The best performance of a PFA can be achieved with a high granularity calorimeter, where it is possible to distinguish between showers created by charged and neutral particles.  Related approaches in other experiments have been described in~\cite{ALEPH,CMS}.

The capability of a PFA  to recover neutral hadron energy in the vicinity of a charged hadron is of crucial importance because mis-assignment of energy would degrade the energy resolution. The mis-assignment of energy between showers is commonly referred to as ``confusion``. This may occur when the event reconstruction algorithm mixes up hits from showers created  by  charged and neutral hadrons as a result of shower overlapping. Another factor which may degrade the energy resolution is the mis-reconstruction of an overlap of a neutral hadron shower and a photon shower. However, to resolve this confusion, in contrast to the case of two hadron showers, energy profiles of electromagnetic showers can be used. In the case of two hadrons producing overlapping showers the task for a PFA becomes more complicated because the energy profiles are less useful and only topological and energy criteria can help to disentangle showers. 

For the International Large Detector (ILD)~\cite{ILD_LOI} proposed for ILC, the PFAs were implemented in a number of reconstruction programs. Among them, the most developed is PandoraPFA \cite{Thomson2009}. It has become a part of the software \cite{ilc_soft} for the ILC and was tested using Monte Carlo (MC) simulated jets. For jet energies of 100--250~GeV, typical for the ILC, the PandoraPFA reconstruction of simulated events in the ILD  concept provides a jet energy resolution of about 3\% which is the goal for the experimental program. It allows the separation of the hadronic decays of $W$ and $Z$ bosons to better than 2.5$\sigma$ for ILC energies and 1.5$\sigma$ for CLIC energies~\cite{Thomson2009}. The implementation of a PFA for CLIC energies is challenging because of the highly boosted jets.

 
The expected performance of a particle flow algorithm at an ILC detector, \textit {e.g.} \cite{Thomson2009}, relies on the ability of the MC simulation to accurately model a number of aspects of hadronic showers. The agreement of MC tools with data allows the optimization of the design of the ILC detectors and therefore is here studied by the CALICE collaboration using test beam data~\cite{had_shower_ecal,had_shower_hcal}. The mis-assignment of reconstructed energy between charged and neutral hadrons in dense jets drives the overall jet energy resolution. It is known that different available physics models give noticeably different predictions for hadron shower shapes, that might be important for resolving the overlapped hadron showers. Moreover, the real detector performance may not be as good as that of the idealized MC model. The main goal of this study is to provide validation of particle flow reconstruction, as implemented in PandoraPFA, using test beam data, and to  compare the result with MC predictions. Such a validation would provide further evidence that the particle flow reconstruction performance for jets obtained in a simulation of the ILC detector concepts is realistic. 

In this paper, the mis-assignment issue is studied by overlaying the hits from two charged pions as observed in the CALICE prototype calorimeter. By shifting the hits from one of the showers in the transverse direction, the effective confusion in the reconstruction can be studied as a function of the shower separation. Test beam data collected at CERN in 2007 by the CALICE detector prototype were used. 

This prototype allows the reconstruction of hadron shower shapes with unprecedented accuracy. It consists of an $\sim 1\lambda_I$ electromagnetic calorimeter (ECAL), an $\sim 5\lambda_I$ hadronic calorimeter (HCAL) and an $\sim 6\lambda_I$ tail catcher and muon tracker (TCMT), where $\lambda_I$ is the nuclear interaction length. The ECAL is a silicon-tungsten sampling calorimeter, made of 30 readout layers with active silicon wafers segmented into diode pads with a size of $1 \times 1$ cm$^2$. The HCAL consists of 38 layers of highly-segmented scintillator plates sandwiched between steel absorber plates. The scintillator segments (tiles) in the zone close to the beam line have dimensions $3 \times 3$ cm$^2$ in the 30 front layers. In the rear and  peripheral regions of the HCAL the  segmentation is coarser. Every tile is read out individually by a silicon photomultiplier (SiPM). The TCMT consists of 16 readout layers interleaved between steel plates. The readout layer consists of twenty $100 \times 5$ cm$^2$ scintillator strips with alternate orientation in odd and even layers, read out by SiPMs.

The detailed description of the complete CALICE setup and first results on hadronic shower reconstruction and analysis can be found in \cite{had_shower_ecal,had_shower_hcal,ecal,hcal}. The CALICE calorimeter prototypes are very similar to the ILD concept. In this study, the longitudinal sampling (layer thicknesses and separations) in the ILD model are modified so as to exactly match the geometry of the CALICE test beam. The cell sizes in the ECAL and HCAL are chosen to be 1$\times$1~cm$^2$ and 3$\times$3~cm$^2$ respectively, corresponding to the CALICE ECAL and the smaller cells of the CALICE HCAL. Even though the prototype HCAL has coarser tiles in its peripheral region and a slightly smaller number of layers than those in the ILD detector concept, for energies up to 30~GeV the structure of hadron showers can be reconstructed with almost the same accuracy as ILD \cite{had_shower_hcal,prot_geom}. To confront test beam data with MC, a GEANT4 \cite{geant4} (version 4.9.2) simulation for two physics lists, LHEP and QGSP\_BERT, was performed using beam profiles corresponding to the data runs. 

\section{Data Selection}
\label{sec_ev_sel}
Single charged pions of 10--30\,GeV were selected in test beam data taken at the SPS (CERN). Backgrounds from electrons and protons were rejected with high efficiency using information from a \u{C}erenkov counter. The remaining background  was also identified and rejected. It is comprised of muons, of events where multiple particles were observed, and of events with low visible energy. Initially the level of the background was 10\% (30\%) for 10~GeV (30~GeV) events.

The selected pion events were subjected to an additional selection procedure based on the energy containment of showers. This is necessary because the HCAL of the detector prototype is not deep enough (ECAL + HCAL~$\sim6\lambda$) to {\em fully} contain every hadronic shower. For ``punch-through`` events, the remaining energy is reconstructed in the TCMT, however this detector is not sufficiently granular to be used in particle flow reconstruction. Hence only pion showers which have more than 95\% of their visible energy contained in the ECAL + HCAL are used for the following analysis.  
Such a selection means that showers which start in the rear of the HCAL are not used. It is worth noting that a shower starting in the rear of the HCAL will be better separated due to the magnetic field in a future detector and hence  the confusion for such a shower will be smaller.

For calibration and reconstruction the CALICE software packages were used \cite{calice_soft}. The detailed description of the calibration and the reconstruction procedure can be found in \cite{ecal,hcal_em}. During the selection procedure only cells containing energy above half that expected from a minimum ionizing particle (0.5\,MIP) were retained in order to reduce the effect of noise. Hereafter such signals will be called hits. The same reconstruction and selection procedures were applied to MC samples after digitization. The analysis is based on about 3300 events.

\section{Overlaying Events}
\subsection{Shifting of Showers}

Charged pion events in the CALICE prototype typically consist of a track-like section followed by a hadronic shower following the primary interaction. To estimate the layer where the primary interaction occurred, an algorithm was designed which, in essence, looks for the point at which the energy per layer and the number of hits increases.\footnote{ In detail, the following algorithm was used.  The moving average $A_i$ of visible energy in MIPs in ten successive layers up to the $i^{\rm th}$ layer and the number of hits in the $i^{\rm th}$ layer $N_i$ are analyzed on a layer-by-layer basis starting from the first ECAL layer. When the conditions $(A_i + A_{i+1}) > (6 + 0.1 \cdot E_{\rm beam}/{\rm GeV})$ MIP and $(N_i + N_{i+1}) > (3.77 + 1.44 \cdot \ln(E_{\rm beam}/{\rm GeV}))$ are satisfied the $i^{\rm th}$ layer is considered to be the primary interaction layer.} Tests on MC samples for the HCAL have shown that the difference between the reconstructed and true primary interaction layer does not exceed 1 layer for 78\% of events and does not exceed 2 layers for more than 90\% of events. 

The part of the event prior to the interaction was termed the {\em primary track}. A neutral hadron nearby to a charged pion was emulated by selecting two charged pion events. In one of the events, all hits up to the identified primary interaction are removed leaving an imitated neutral hadron shower. In what follows, we will call the energy of this shower the {\it neutral hadron shower} energy. This neutral hadron has a slightly reduced energy compared to the energy of the original beam particle since the energy lost up to the first interaction is not considered, but for simplicity such particles are always referred to with the energy of the original particle in the following. The hit positions of the neutral hadron shower are shifted in the transverse direction by between 5 cm and 30 cm and are then superimposed with the hits from the other selected charged pion. Since the shifting procedure requires determination of the shower axis position, the pion entry point into the calorimeter (primary track coordinates) was identified for each event. 

\begin{figure}[ht]
\includegraphics[width=.33\textwidth]{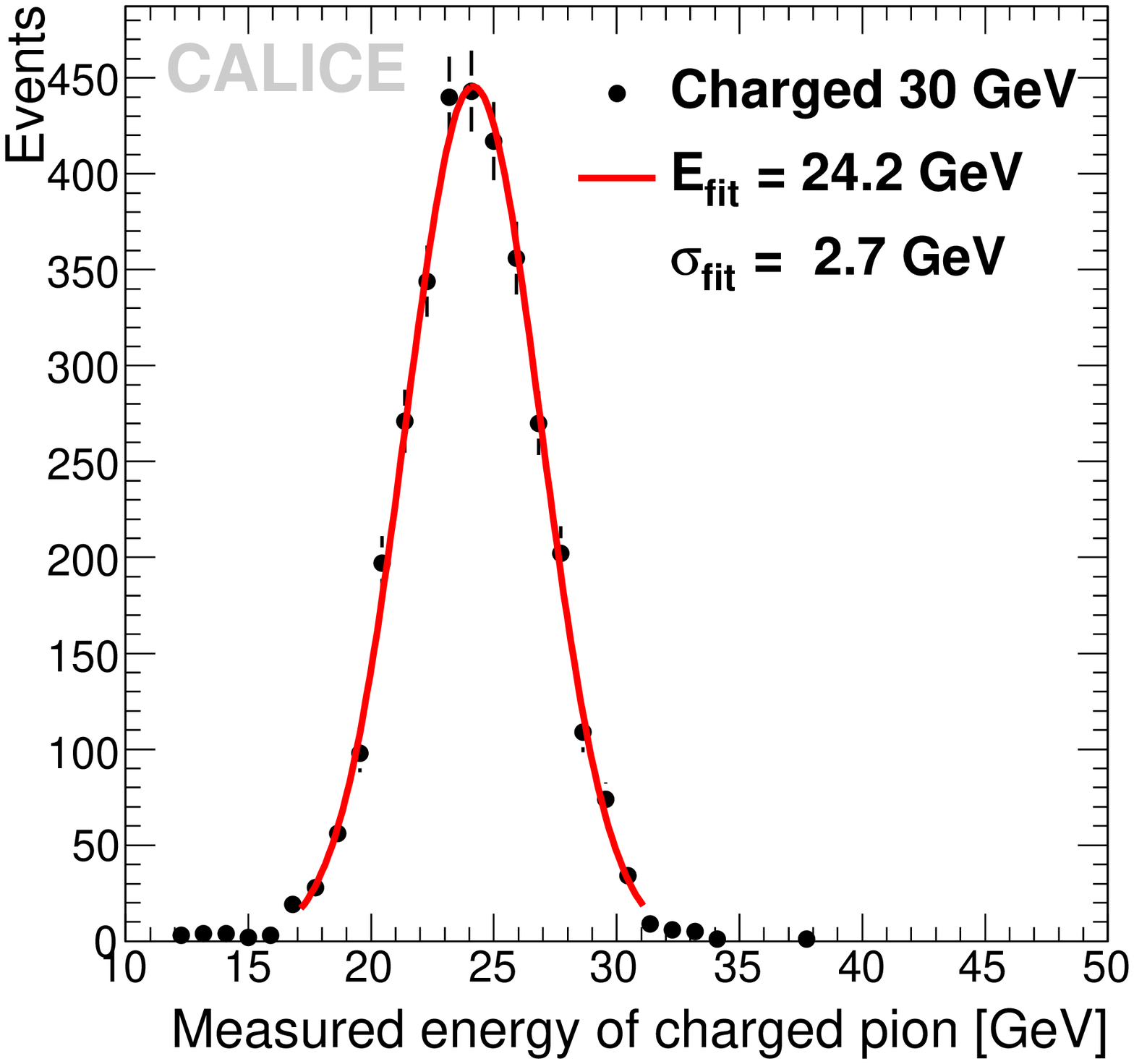}
\includegraphics[width=.33\textwidth]{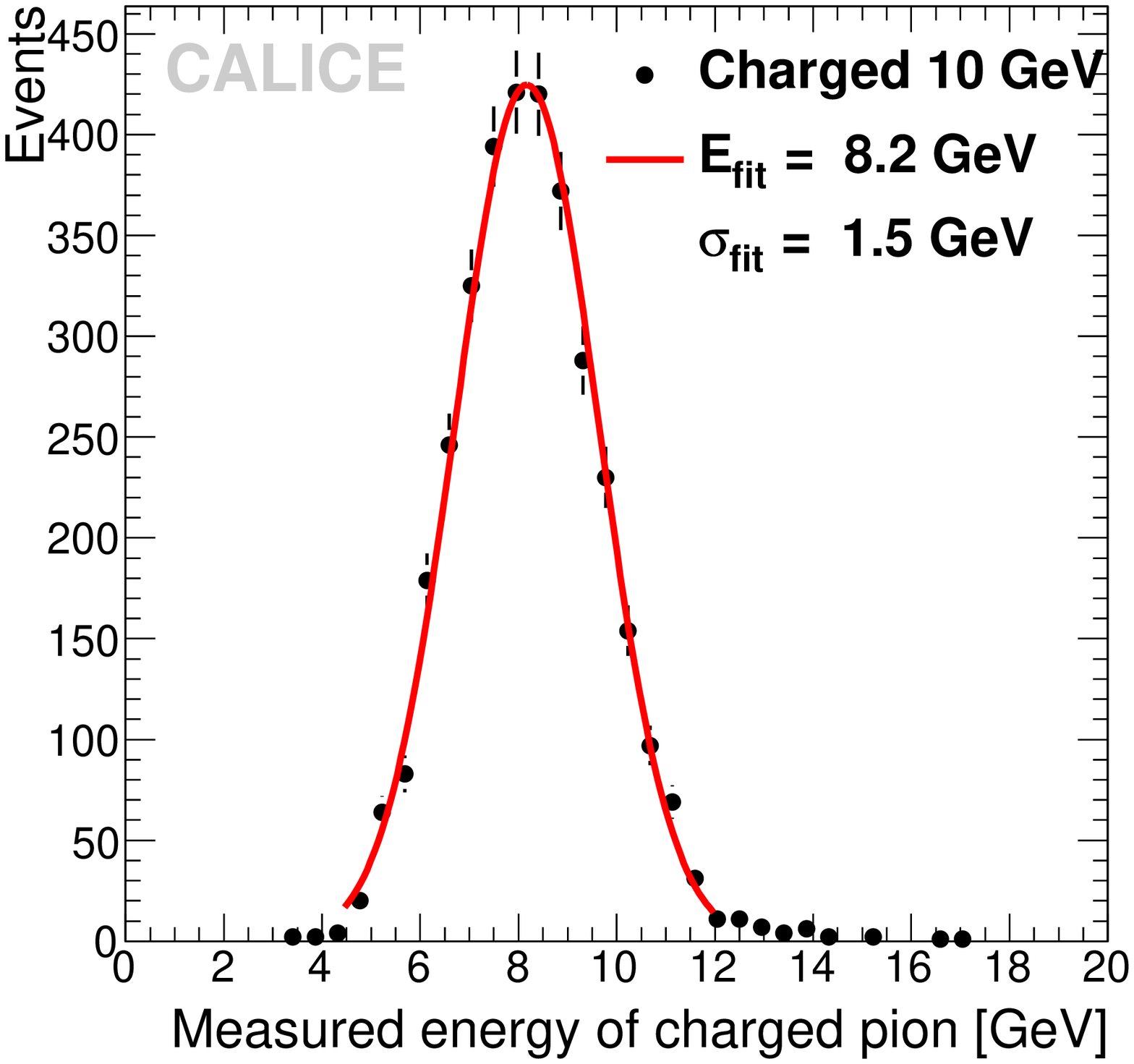}
\includegraphics[width=.33\textwidth]{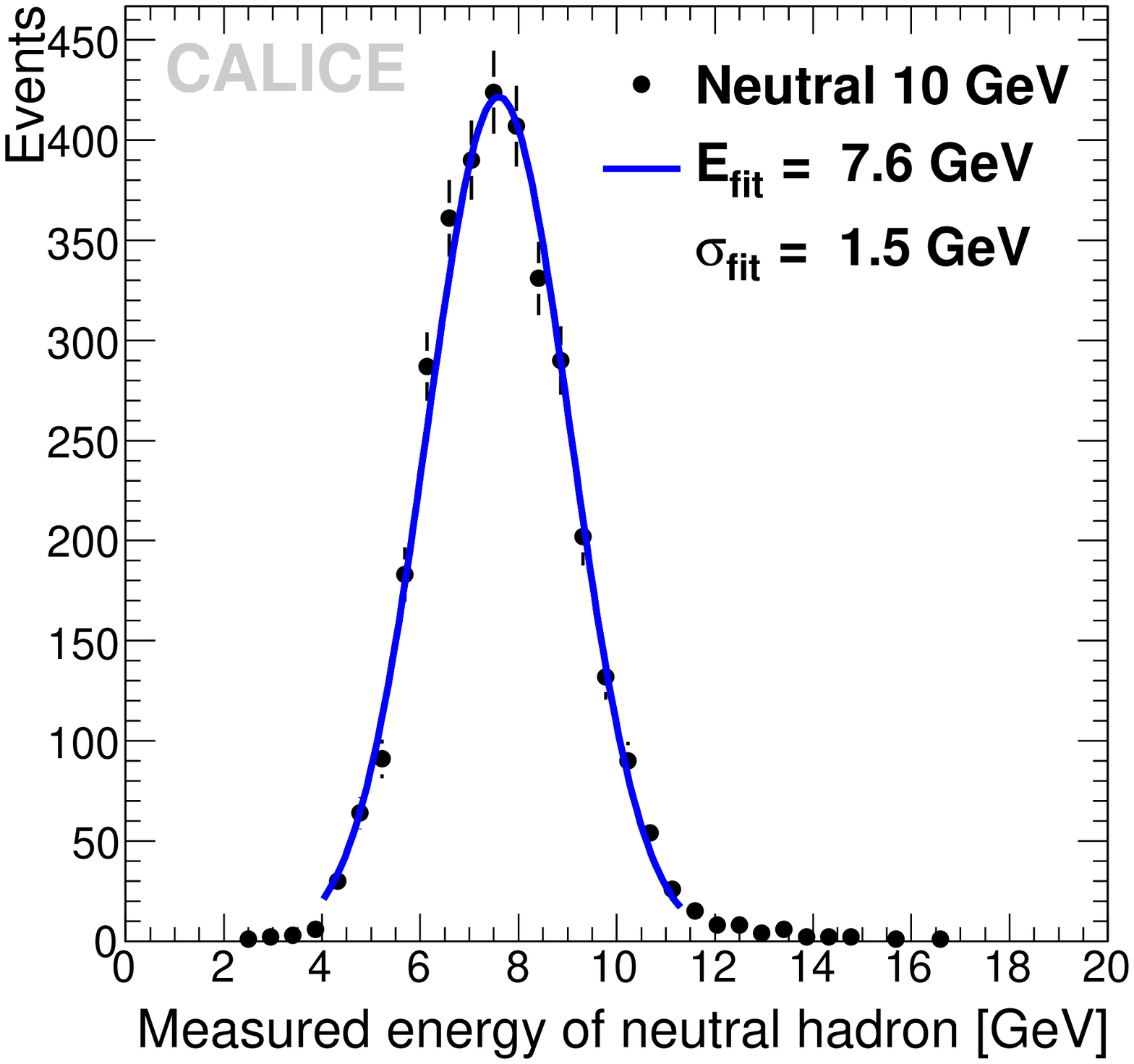}
\caption{Energy distributions for 30~GeV charged (left), 10~GeV charged (middle) and 10~GeV ``neutral'' (right) hadron events prepared from data runs for mixing of two showers. Solid lines correspond to Gaussian fits with mean $E_{fit}$ and sigma $\sigma_{fit}$ indicated in the legend.}
\label{ener}
\end{figure}

Figure~\ref{ener} shows the energy distributions for the 30~GeV charged (left), 10~GeV charged (middle), and 10~GeV ``neutral'' (right) hadron events. These energies were chosen for the following analysis as being representative of a 100~GeV jet \cite{Wigmans2002,Green1990}. The error bars are purely statistical.
Throughout this paper, the calibration of the CALICE prototypes was defined so as to reconstruct the energy for electrons. Therefore the distributions for hadrons peak at lower energies than the beam energy, by approximately 20\% (reflecting the $\pi/e$ response ratio). In addition, the energy of the emulated ``neutral'' hadrons peaks lower than for the charged pions because the energy deposited by the incoming particle is discarded.

\subsection{PandoraPFA Adjustment}
The PandoraPFA  represents the state of the art in high granularity Particle Flow Reconstruction. The original version of PandoraPFA assumed a collider detector geometry of a central tracker and barrel and end-cap calorimeters.\footnote {The latest version of PandoraPFA which makes no assumptions about the detector geometry has only recently been released \cite{PandoraNew}.} Due to the limitations of the existing software it was necessary to map the CALICE events onto a collider detector geometry. This was relatively simple since the CALICE prototype has a very similar geometry to the ILD detector at the ILC.

The events with two overlaid showers were mapped to the ILD structure with the dimensions of ECAL cells and with the number and thicknesses of layers and absorbers equal to those in CALICE calorimeter prototypes. The ILD detector is an octahedral barrel with two endcaps (see \cite{ILD_LOI}). The CALICE prototype hits were put in the top octant of the barrel, layer by layer. Thus the CALICE beam became directed vertically up along the \textit{y} axis in the ILD geometry. 
Since the tiles of the ILD HCAL have transverse dimensions $3 \times 3$ cm$^2$, there is a difference in the transverse cell sizes between the prototype and ILD HCAL in peripheral regions of showers. The coarse cells in the prototype HCAL sample the lateral and longitudinal tails of the showers, where the particle density is low.  Therefore, rather than subdividing the energy of a CALICE hit between several cells in the ILD calorimeter, instead the energy was simply placed in the ILD cell whose centre lay closest to the centre of the CALICE cell.
The original distance between hits was thus preserved except for the border between the coarse and small cells. This slightly affects the shower topology in a way which complicates the task for PandoraPFA to resolve two showers, making our conclusions about PandoraPFA performance rather conservative.  

To make the energy comparison fair, the sum of the first and the second shower energy measured by the prototype should stay equal to the full energy written in the  two hadron event. For this reason, signals from two hits in the same tile were simply added together in the process of shower merging. The possibility that the sum of two signals below the 0.5~MIP threshold exceeds the threshold after shower merging is ignored since such signals contribute about 0.1\% of the energy of overlaid  showers even when there is a small distance between them. Together with hits, the energy  of the two showers measured by the CALICE prototype was written out and passed to PandoraPFA for comparison with the recovered energy of the showers. An identical procedure was applied to MC simulated showers. Inside the program, the CALICE calibration coefficients were used. The energy of showers was left at the calibration (electromagnetic) scale. The reclustering algorithm of PandoraPFA assumes knowledge of the energies of the charged particles based on tracking.  In our case, this is replaced by the known beam energy, scaled by an estimation of the $\pi/e$ ratio based on fits to distributions of the original measured energy such as those in figure~\ref{ener}. To account for small scale differences between beam data 
and simulations with different physics lists, the $\pi/e$ ratio is determined for each of the data sets separately.

As the CALICE prototype was tested without a magnetic field, the PandoraPFA processor  has been adjusted and simplified for this study. Modifications were made to the way in which the charged hadron track parameters are treated. To calculate the energy of the track and its entrance point position and direction, the assumed TPC track helix made from a fit to the TPC hits was replaced by a simple straight track projection which intersects the calorimeter barrel inner surface at zero \textit{xz} position with normal incidence and has definite energy given by the scaled beam energy as explained above. Subsequent calculations of a distance between the helical track extrapolation and shower hits or clusters  were replaced by a calculation of the distance from the extrapolation of the straight track.  In the presence of a magnetic field, even though the hadron shower gets a little smearing, its end appears to be further from the jet axis than the shower beginning. Thus, the magnetic field makes it easier for PandoraPFA to separate showers. Therefore our analysis gives a conservative estimate for the PandoraPFA performance. 

A number of the methods in the PandoraPFA algorithm, such as kink track cluster association, primary photon recovering, and multi-track cluster association splitting, are not appropriate for this analysis and are not run. The assignment of unused isolated hits and small ($< 10$ hits) clusters  is done proportionally to the estimated energies of the charged and neutral hadrons. To calculate this proportion, the energy of the neutral hadron is taken equal to the difference between the summed energy of both hadrons measured in the calorimeter, and the mean energy of the charged hadron. Such an assignment gives actually zero mean difference between recovered and measured energies for a 10~GeV neutral hadron at large distances from a 10~GeV charged hadron, see section \ref{sec_recovering}.

\section{Recovering of Showers}
\label{sec_recovering}

The PandoraPFA is a very sophisticated multi-stage program which includes stages of clustering, reclustering  and the removal of neutral fragments. The clustering algorithm is a cone-based procedure followed by topological merging of clusters. Reclustering is an iterative algorithm aiming to make consistent the cluster energy and the information of the associated track. Finally, both topological and energy criteria are utilized to merge fragments of charged clusters with parent clusters. In all, there are several tens of algorithms in the program, each of which corrects deficiencies of previous stages, thereby improving the reconstruction of showers. Due to the  overlapping  showers in the calorimeter, the energy recovered by PandoraPFA for each of the showers is not always accurate.
 
\begin{figure}[ht]
\includegraphics[width=.5\textwidth]{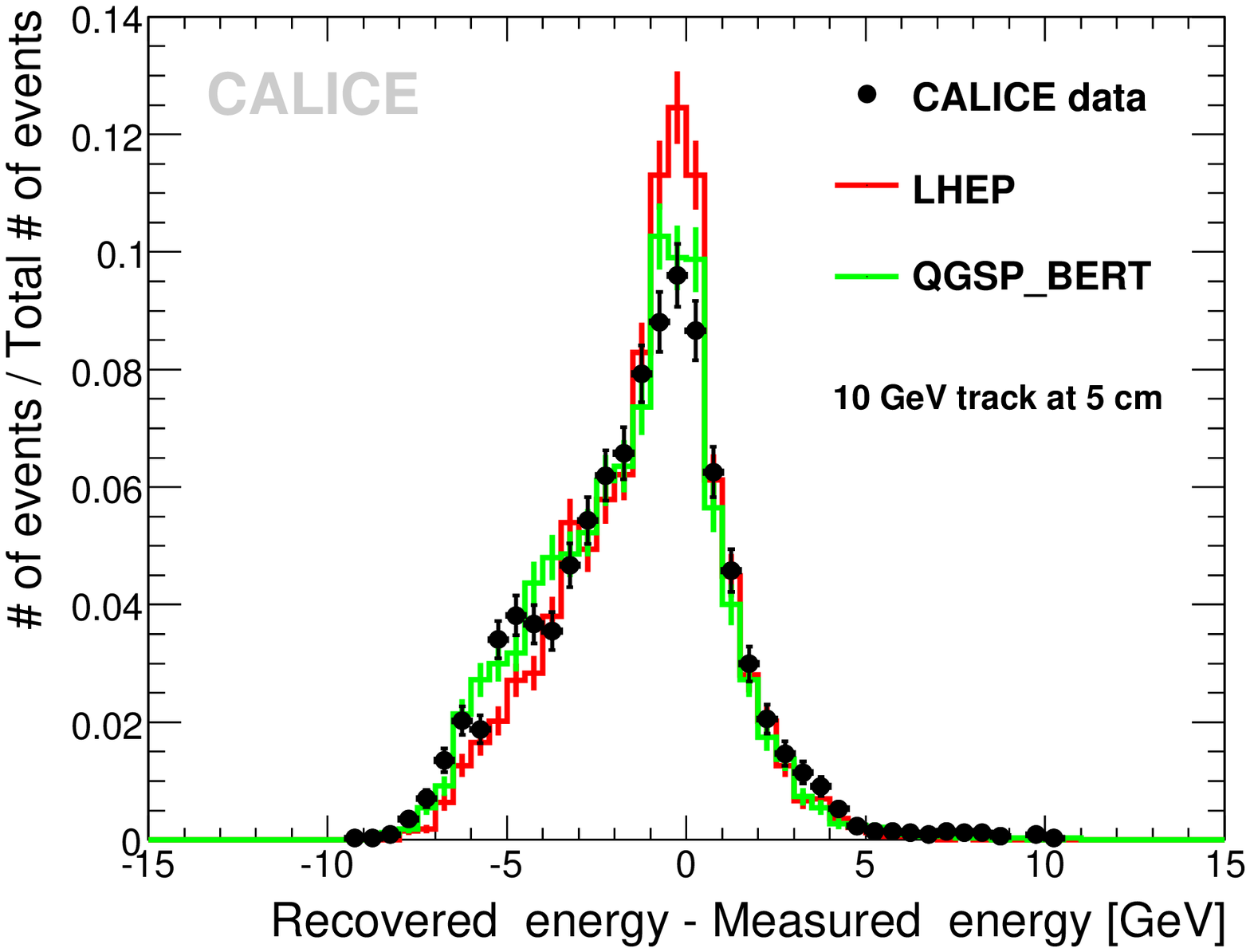}
\includegraphics[width=.5\textwidth]{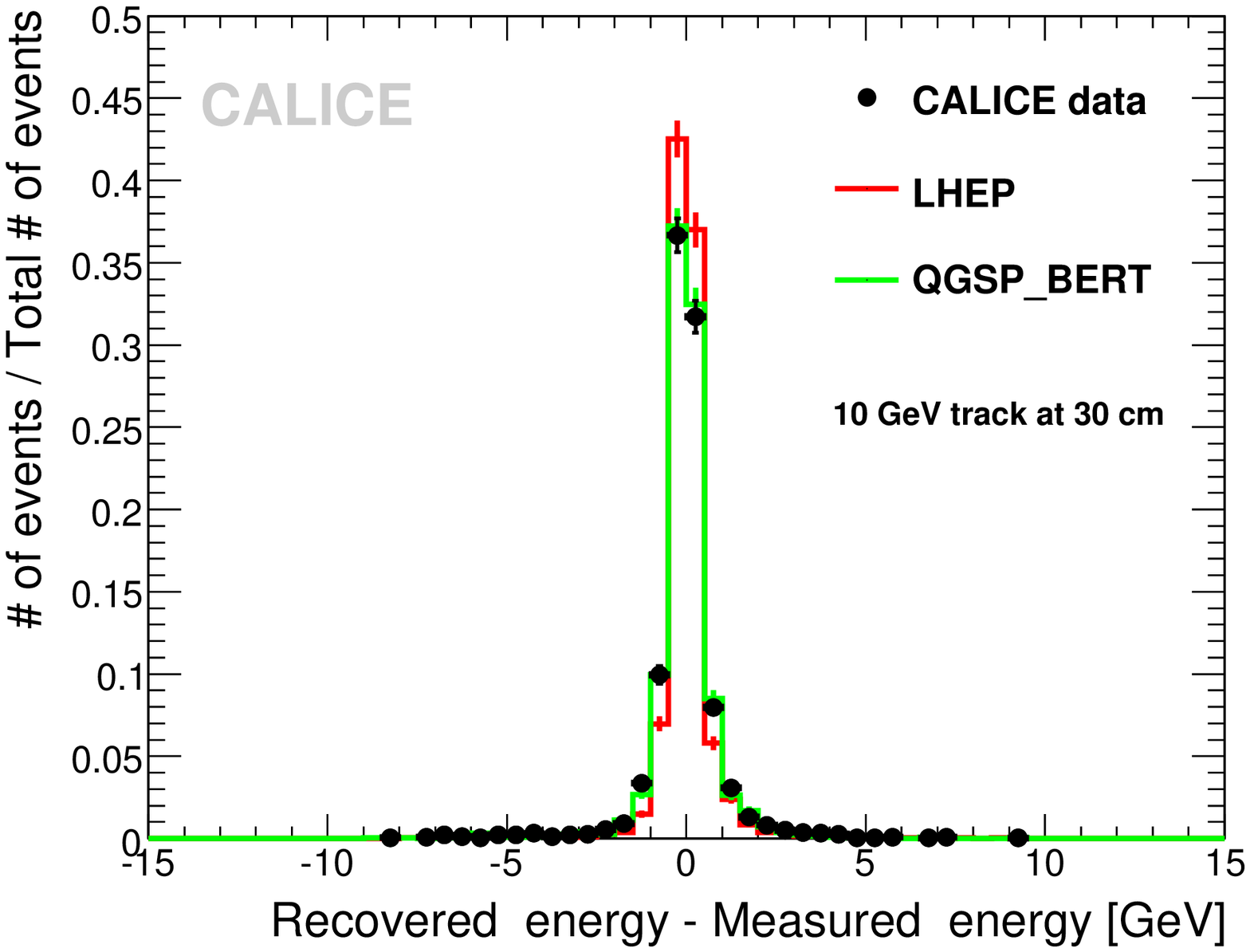}
\newline
\includegraphics[width=.5\textwidth]{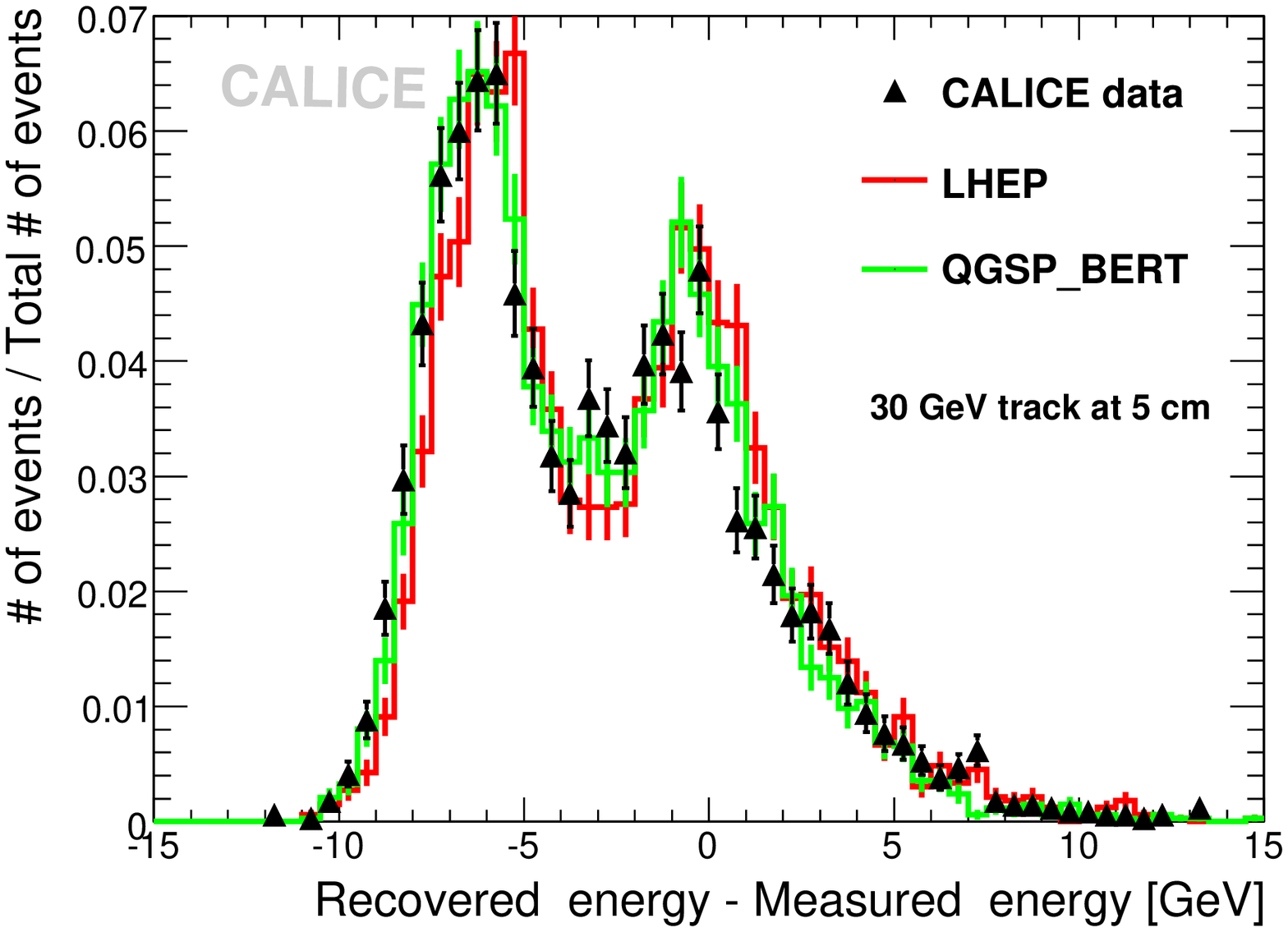}
\includegraphics[width=.5\textwidth]{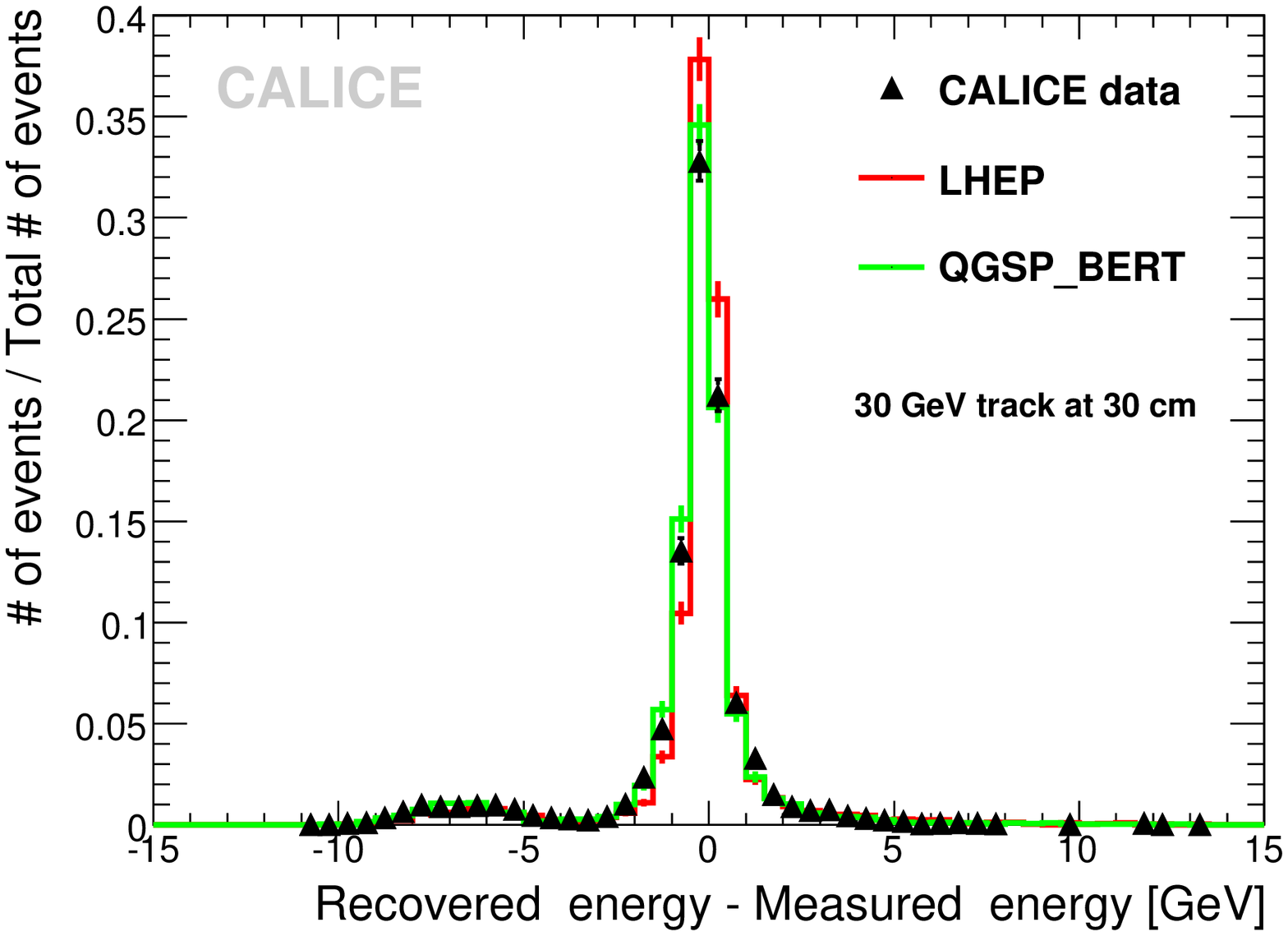}
\caption{ Difference between the recovered energy and the measured energy
for the 10~GeV neutral hadron at 5 cm (left) and at 30 cm (right) from the 10~GeV (top) and 30~GeV (bottom) charged hadrons. Data (black) are compared to MC predictions for LHEP (red) and QGSP\_BERT (green) physics lists.} \label{dif_d}
\end{figure}

In this study, since the neutral hadron energy is known from the original (single particle) calorimetric measurement, this can be compared to the reconstructed energy from PandoraPFA to obtain an estimate of the level of confusion. In making this comparison, note that the original calorimetric measurement for the neutral hadron is lower than the appropriate beam energy (figure~\ref{ener} right), since the calibration is to electromagnetic energy, and the ionization energy deposited by the incoming track is lost. Figure~\ref{dif_d} shows the difference between the energy recovered by PandoraPFA and the original measured energy for a 10 GeV neutral hadron shower in the vicinity of a charged pion shower for two distances between them and for two charged pion energies.  These distributions can be interpreted in terms of the confusion introduced by the pattern recognition. The maximum confusion takes place between a high energy charged hadron and a low energy neutral hadron (see bottom left plot in figure~\ref{dif_d}). The confusion is particularly large for events in which, due to intrinsic shower fluctuations, the difference between the measured charged hadron energy and the beam energy is comparable with  the neutral  hadron energy. This results in a peak around $-7$~GeV for a 30~GeV charged and a 10~GeV neutral hadron (see figure~\ref{dif_d}). At large distances this confusion largely vanishes. For a 10~GeV charged hadron, the neutral hadron energy reconstruction is considerably better (see top plots in figure~\ref{dif_d}).

\begin{figure}[ht]
\centering
\includegraphics[width=.8\textwidth]{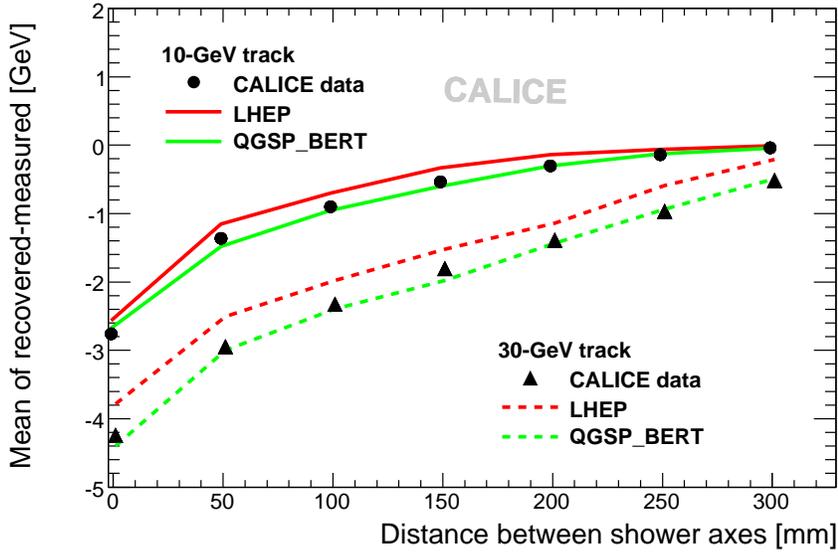}
\caption{ Mean difference between the recovered energy and the measured energy for 10~GeV neutral hadrons vs. the distance from 10~GeV (circles and continuous lines) charged hadrons and  30~GeV (triangles and dashed lines) charged hadrons.} \label{mean_vs_r}
\end{figure}

\begin{figure}[ht]
\includegraphics[width=.5\textwidth]{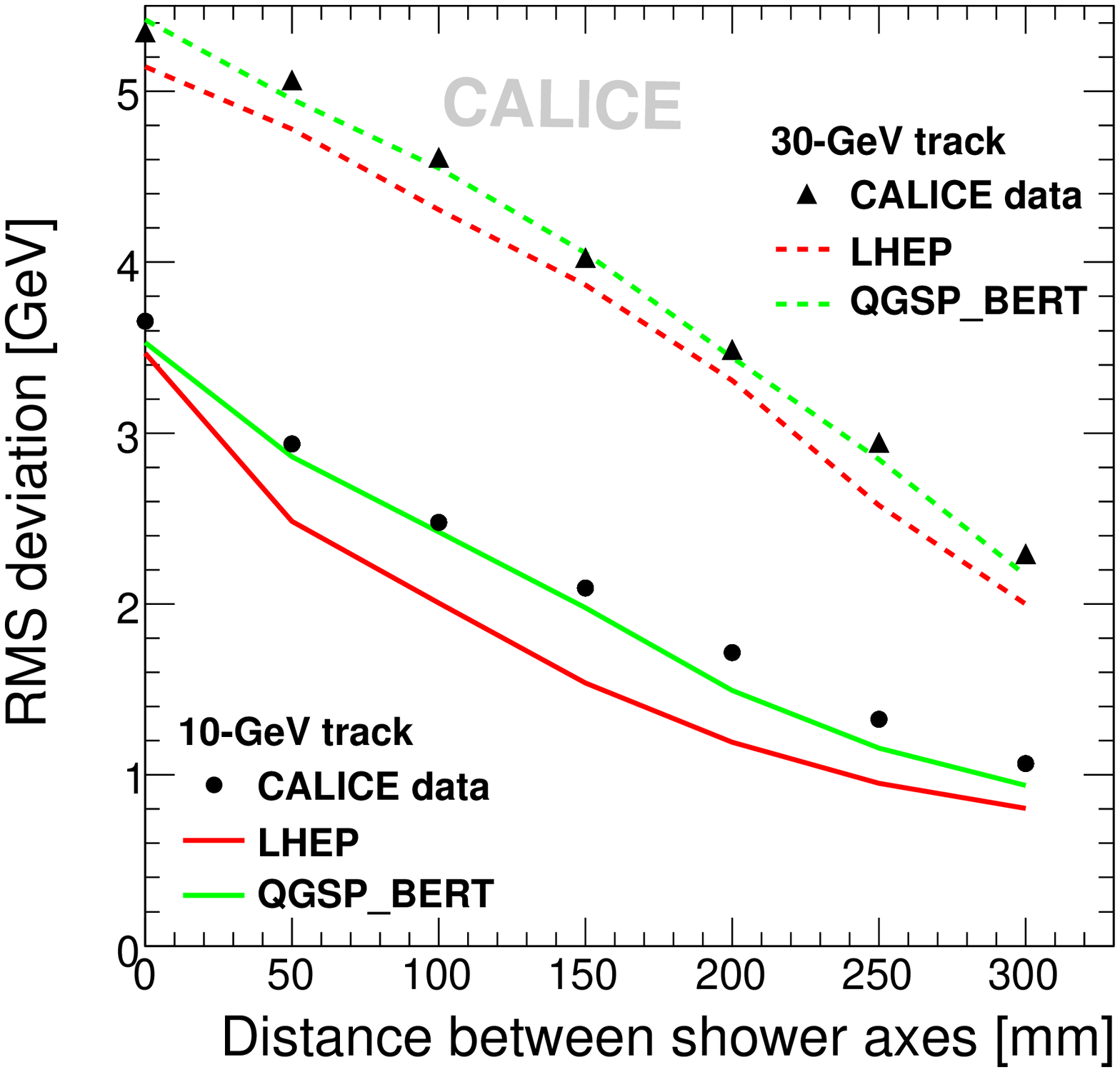}
\includegraphics[width=.5\textwidth]{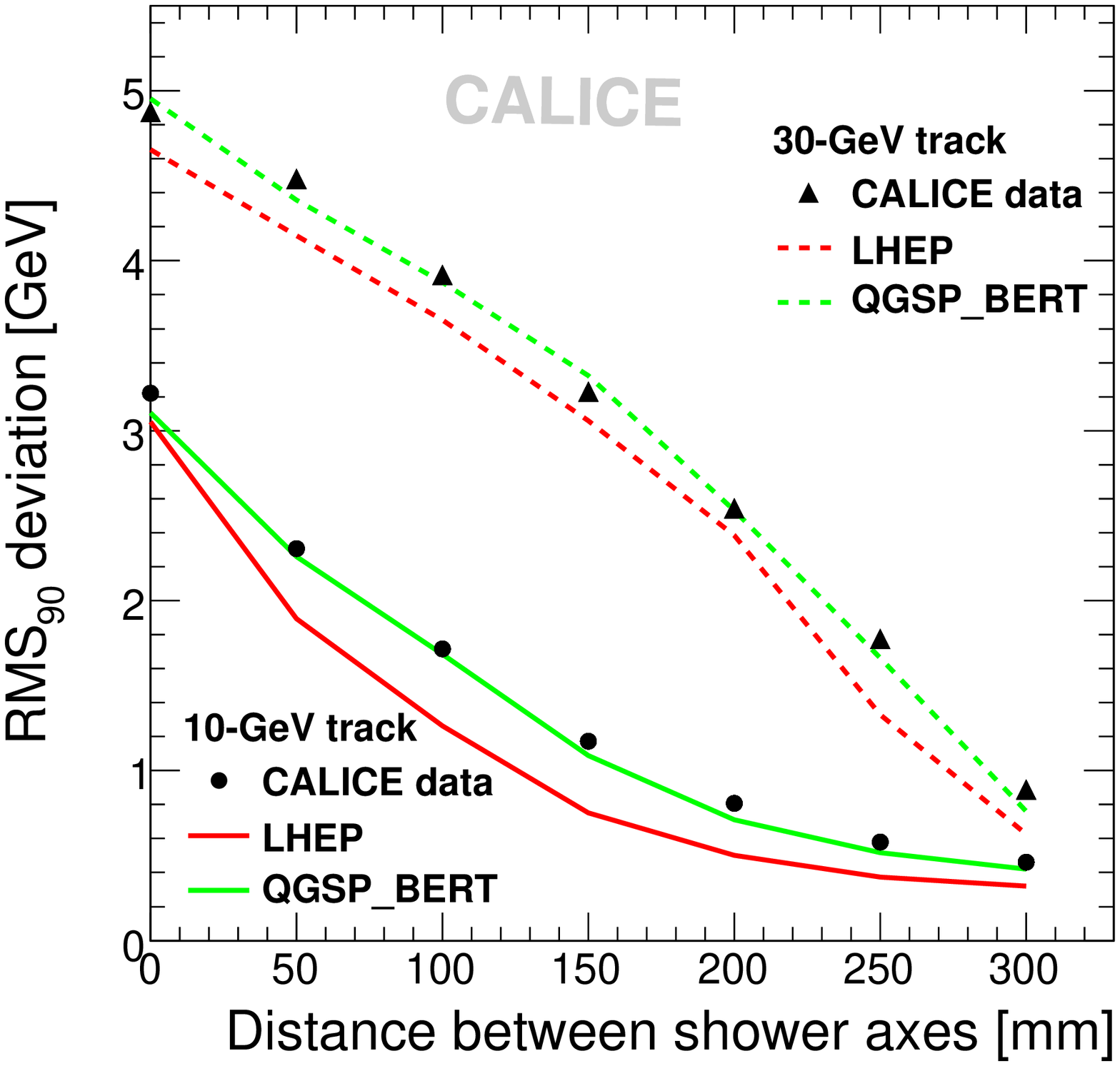}
\caption{ $RMS$ (left) and $RMS_{90}$ (right) deviations of the recovered energy of neutral 10~GeV
hadrons from its measured energy vs. the distance from 
charged 10~GeV (circles and continuous lines) and 30~GeV (triangles and dashed lines)
hadrons for beam data (black) and for Monte Carlo simulated data, for
both LHEP (red) and QGSP\_BERT (green) physics lists.} 
\label{rms_0_vs_r}
\end{figure}

>From the plots shown in figure~\ref{dif_d} the mean value of the difference between recovered energy and original measured energy of a neutral hadron can be extracted.  At small distances  between particles where  shower overlap is considerable, the mean energy of the neutral hadron recovered by PandoraPFA is typically lower than the corresponding energy measured in the calorimeter prototype (see figure~\ref{mean_vs_r}). Due to a successful performance of the reclustering  algorithm, even at zero distance, PandoraPFA recovers the neutral cluster energy correctly in a large fraction of events. The confusion naturally depends on the transverse size (radius) of showers and their internal structure. Therefore, the LHEP based simulation which gives narrower and more compact showers, predicts smaller confusion than is seen in data, while the simulations based on the QGSP\_BERT physics list describe the data better, see figure~\ref{mean_vs_r} (and also figure~\ref{rms_0_vs_r}, described below).

The second characteristic used to estimate the confusion error is the root mean square ($RMS$) deviation. However, to avoid the over-emphasizing of the distribution tails, the $RMS_{90}$ value is  used. It is defined as the RMS deviation of the recovered energy from the energy measured in the calorimeter prototype in the central region of the distribution which contains 90\% of the events (see e.g. \cite{Thomson2009}). The RMS deviation of the recovered energy of a neutral hadron from its measured energy can be interpreted as a confusion error. It is particularly large for the 30~GeV charged and overlapping 10~GeV neutral hadrons, see figure~\ref{rms_0_vs_r}. However, this does not affect the jet energy reconstruction accuracy at the ILC too much because the probability to find a 30~GeV charged particle in a 100~GeV jet is relatively low \cite{Green1990,Knowles1997}.

\begin{figure}[ht]
\includegraphics[width=.5\textwidth]{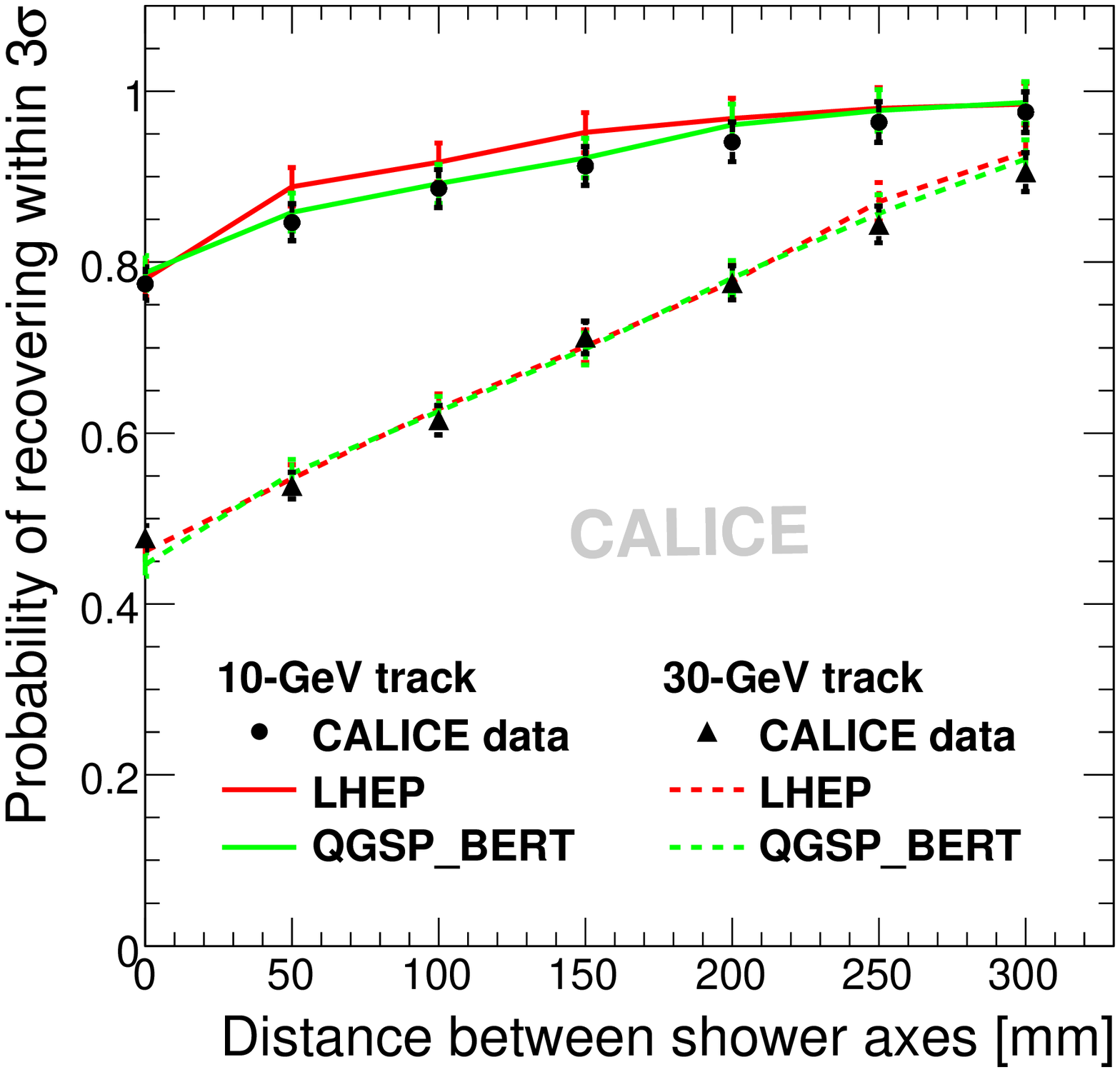}
\includegraphics[width=.5\textwidth]{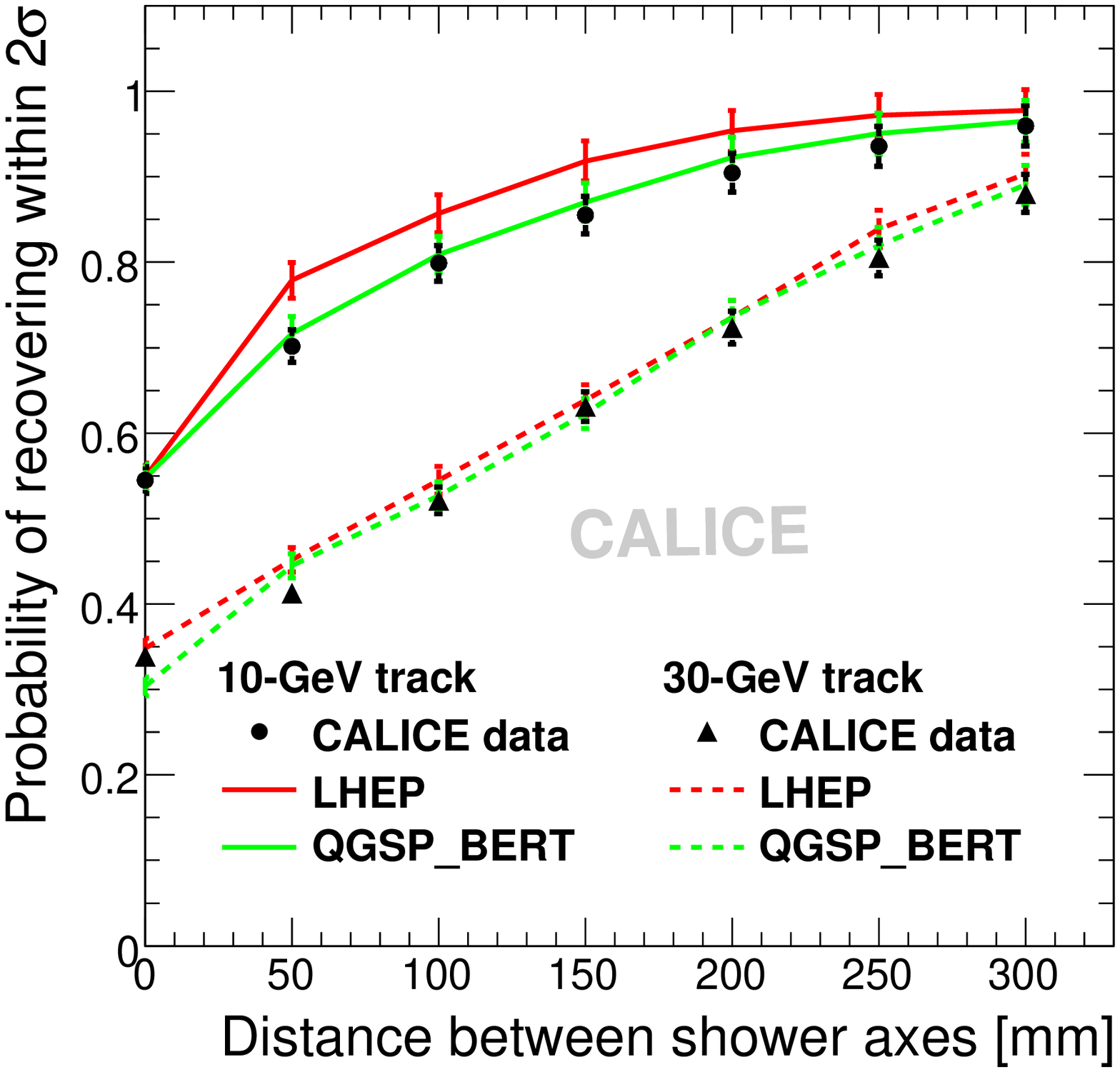}
\caption{ Probability of  neutral 10~GeV hadrons energy recovering within
3 (left) and 2 (right) standard deviations from its real energy vs. the distance from charged 10~GeV
(circles and continuous lines) and 30~GeV (triangles and dashed lines) hadrons for beam data
(black) and for Monte Carlo simulated data, for both LHEP (red) and
QGSP\_BERT (green) physics lists.} 
\label{eff_dist}
\end{figure}

Figure~\ref{eff_dist} shows the probability of recovering of the 10~GeV neutral hadron energy  within 2 and 3 standard deviations from its real energy at different distances from 10~GeV and 30~GeV charged hadrons. For the beam data neutral hadron we take the standard deviation equal to \newline $0.55\sqrt{10\times0.82-0.6}$~GeV. Here the coefficients  0.55 and 0.82 are estimations of the stochastic term  coefficient and the $\pi/e$ ratio of the calorimeter prototype respectively, based on fits to distributions of the original measured energy such as those in figure~\ref{ener}.  The 0.6~GeV is the average primary track loss for the imitated neutral shower, estimated from the difference between the mean value of the energy distributions before and after the removal of the primary track. For the MC simulated neutral hadrons the standard deviation is calculated  in the same manner,  but using estimations based on fits to the appropriate distributions. 

\begin{figure}[ht]
\includegraphics[width=.5\textwidth]{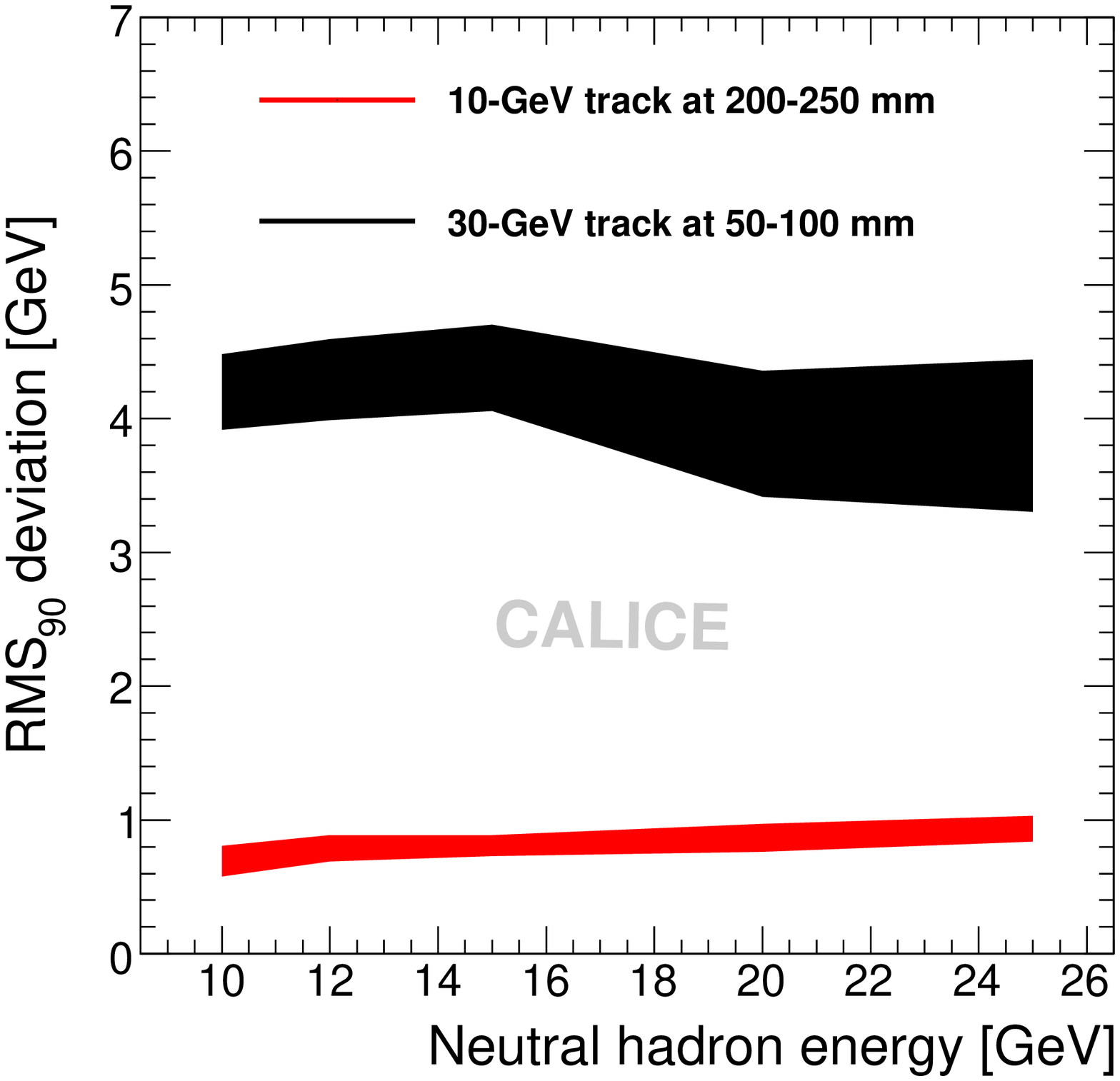}
\includegraphics[width=.5\textwidth]{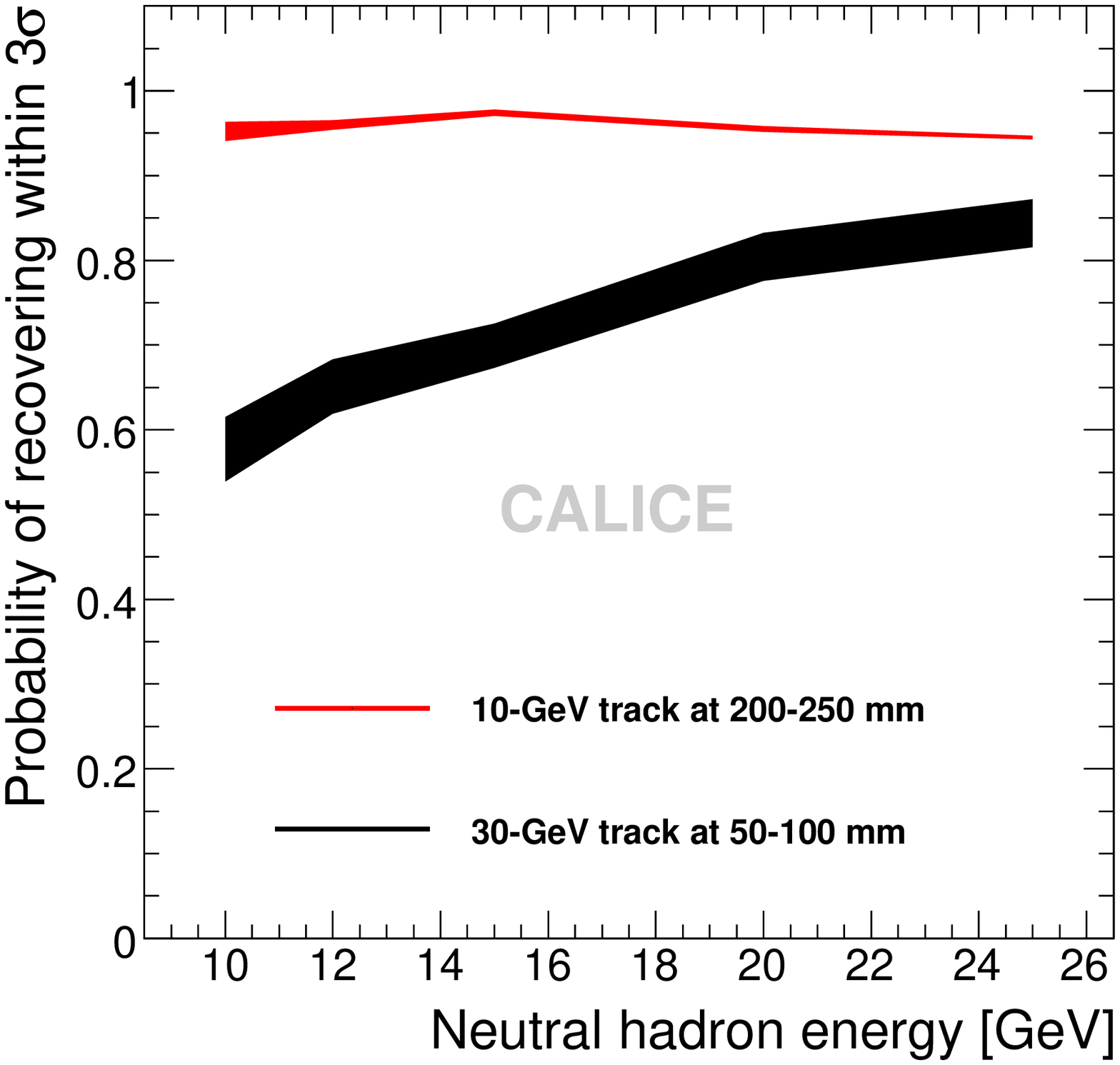}
\caption{ $RMS_{90}$ deviation of the recovered energy of neutral hadrons from their measured
energy (left) and the probability of the neutral hadron energy recovering
within 3 standard deviations (right) vs. the neutral hadron energy in the
vicinity of 10~GeV charged hadrons (red) and 30~GeV charged
hadrons (black), see text for more details.}
\label{rms0_vs_neutral}
\end{figure}

If the charged hadron is situated in the vicinity of a neutral hadron with similar or higher energy, the confusion is typically less than in the reversed situation. In figure~\ref{rms0_vs_neutral} we use the test beam data to estimate how the confusion depends on the energy of the neutral hadron.  In jets in a full detector such as ILD, the charged particles will tend to be separated from the  neutrals by the magnetic field. Therefore, in this figure the charged  hadron is placed at a distance typical of its deflection in a 4\,T magnetic field in the ILD geometry.
The $RMS_{90}$ deviation of the recovered neutral hadron energy from its measured energy does not depend significantly on the neutral hadron energy (see left plot in figure~\ref{rms0_vs_neutral}). The relative confusion is large for small neutral hadron energy. This results in a smaller probability of neutral hadron energy recovery for small neutral hadron energy (see right plot in figure~\ref{rms0_vs_neutral}).

\section{Summary}
To test the particle flow algorithm, PandoraPFA, we have mapped pairs of CALICE test beam events, shifted by the definite distances from each other, onto the ILD geometry. Then we modified the treatment of tracks in the PandoraPFA processor for the case of straight tracks. In this study we have investigated the hadron energy range typical for a 100~GeV jet. For jet fragment energies from 10~GeV to 30~GeV we estimated the confusion error for the recovered neutral hadron energy caused by the overlapping of showers. 

We have confronted our result for test beam data with the result of Monte Carlo simulations for LHEP and QGSP\_BERT physics lists. The results for the data and MC are in a good agreement. This fact together with the successful PandoraPFA performance for simulated jets \cite{Thomson2009} allows us to consider the PandoraPFA program as a good reconstruction tool for a full-size experiment. Our results for the confusion are overestimated; in a full-size experiment the program would give smaller confusion. In particular, the fact that the prototype HCAL does not have a fixed tile size complicates the clustering procedure. Additionally, we underestimate the separation of showers towards the end of the calorimeter because, unlike a full detector, our testbeam apparatus has no magnetic field.

The agreement between the PandoraPFA performance achieved with real calorimeter prototype data and with the MC simulation demonstrates that the extrapolation to the complete detector is reliable. No hidden imperfections in the real data (imperfect calibration, non-uniformity of tile response, cross talk between tiles, dead or noisy channels) which could degrade the PFA performance were found. In particular, this conclusion is in agreement with the results of a study of the impact of tile non-uniformity reported in  \cite{Angela}. We find in our study that the QGSP\_BERT physics list gives a better description of test beam data than does LHEP.


\section{Acknowledgments} 

We would like to thank the technicians and the engineers who
contributed to the design and construction of the prototypes. CALICE conducts 
test beams at CERN, DESY and FNAL and we gratefully acknowledge the managements of these laboratories 
for their support and hospitality, and their accelerator staff for the reliable and efficient
beam operation. 
This work was supported within the 'Quarks and Leptons' programme of the CNRS/IN2P3, France;
Bundesministerium f\"{u}r Bildung und Forschung, grant
no.\ 05HS6VH1, Germany;
by the DFG cluster of excellence `Origin and Structure of the
Universe' of Germany; 
by the Helmholtz-Nachwuchsgruppen grant VH-NG-206;
by the Alexander von Humboldt Foundation (Research Award IV, RUS1066839 GSA);
by joint Helmholtz Foundation and RFBR grant HRJRG-002, SC Rosatom;
by Russian Grants SS-3270.2010.2, RFBR07-02-92281,
RFBR08-02-12100-OF, RFBR09-02-91321, by the Russian Ministry for Education and Science and
by Russian National Educational Center grant 02.740.11.0239;
by MICINN and CPAN, Spain;
by the US Department of Energy and the US National Science Foundation;
by the Ministry of Education, Youth and Sports of the Czech Republic
under the projects AV0 Z3407391, AV0 Z10100502, LC527 and LA09042 and by the
Grant Agency of the Czech Republic under the project 202/05/0653;  
and by the Science and Technology Facilities Council, UK.

\end{document}